\documentclass[conference,10pt]{IEEEtran}
\usepackage{amsfonts}
\usepackage{amssymb}
\usepackage{setspace}
\usepackage[latin1]{inputenc}
\usepackage{graphicx}
\usepackage{amsthm}
\usepackage{color}
\usepackage{cite}
\usepackage{bm}
\usepackage{epsfig,psfrag}
\usepackage{tabularx}
\usepackage{tikz}
\usepackage{pst-node}
\usepackage{acronym}
\usepackage[yyyymmdd,hhmmss]{datetime}
\usepackage{notation}
\usetikzlibrary{arrows,backgrounds,calc,positioning,shapes,shadows}
\usepackage{gensymb}
\usepackage{tikz}

\usepackage{amsmath}
\usepackage{graphicx}			% Include figure files
\usepackage{epstopdf}
\usepackage{xcolor}
\usepackage{epsfig,psfrag}
\usepackage{notation}

\DeclareMathAlphabet{\mathpzc}{OT1}{pzc}{m}{it}

\newcommand{\ist}{\hspace*{.3mm}}
\newcommand{\rmv}{\hspace*{-.3mm}}

\newcommand{\nn}{\nonumber}

\newcommand{\trans}{^\mathrm{T}}

\definecolor{greenNew}{RGB}{0,170,0}
\definecolor{grayNew}{RGB}{50,50,50}

\allowdisplaybreaks

\begin{document}
\title{A Probabilistic Focalization Approach\\ for Single Receiver Underwater Localization}

\author{\IEEEauthorblockN{Luisa Watkins\IEEEauthorrefmark{1}, Pietro Stinco\IEEEauthorrefmark{2}, Alessandra Tesei\IEEEauthorrefmark{2},
and Florian Meyer\IEEEauthorrefmark{1}}\vspace{.8mm}
\IEEEauthorblockA{\IEEEauthorrefmark{1}Scripps Institution of Oceanography and Department of Electrical and Computer Engineering\\ University of California San Diego, La Jolla, CA (\{l1watkins,flmeyer\}@ucsd.edu)\vspace{1.5mm}}
\IEEEauthorblockA{\IEEEauthorrefmark{2}NATO STO Centre for Maritime Research and Experimentation, La Spezia, Italy\\ (\{pietro.stinco, alessandra.tesei\}@cmre.nato.int) }
 \vspace*{-7mm}}

\maketitle

\begin{abstract}
We introduce a Bayesian estimation approach for the passive localization of an acoustic source in shallow water using a single mobile receiver. The proposed probabilistic focalization method estimates the time-varying source location in the presence of measurement-origin uncertainty. In particular, probabilistic data association is performed to match time-differences-of-arrival (TDOA) observations extracted from the acoustic signal to TDOAs predictions provided by the statistical model. The performance of our approach is evaluated using real acoustic data recorded by a single mobile receiver. \vspace{3mm}\end{abstract}


\acresetall
%%%%%%%%%%%%%%%%%%%%%%%%%%%%%%%%%%%%%%%%%%%%%%%%%%%%%%%%%%%
\section{Introduction} \label{sec:intro}

Underwater passive acoustic source localization and tracking aims to estimate a source's position in the water column using properties of the acoustic source and knowledge of the environment.  In shallow water, it is unlikely that only one acoustic signal is propagated from the source to the receiver, but rather, there are multiple propagation paths that the signals may follow, resulting in a received signal that is the superposition of several multipath components. 
Passively listening to acoustic signals and exploiting multipath propagation to localize an acoustic source is a key task in a variety of applications, including geoacoustic inversion, marine biology, and maritime surveillance. Source localization methods that only require a single receiver can be advantageous compared to methods that rely on multiple receivers due to their ability to be used on small platforms with limited, or even without, the capability of towing a receiver array. When such a small platform is used and data is collected from a single receiver, advanced signal processing methods are necessary to extract the individual multipath components \cite{oppenheim2004frequency}. The extracted components are used as measurements and are subject to measurement-origin uncertainty \cite{BarWilTia:B11}, i.e., it is not clear which component was generated by which propagation path, and the presence of missed paths and false measurements further complicates source localization.

\subsection{State-of-the-art Methods}

A scientific approach to source localization is \textit{matched field processing (MFP)}, which uses properties of the environment, such as the \textit{sound speed profile (SSP)} and water depth, as well as the geometry of the used receivers, to compute an estimate of the source's position in both range and depth \cite{baggeroer1993overview}. MFP aims at maximizing the power of the source's signal over an entire pressure field using a grid of test points. In order to perform estimation accurately, the ocean's environment needs to be properly modeled. While typically applied to an array of hydrophones, recent works have continued to develop MFP for single receiver source localization. 
One study combines an MFP approach with deep neural networks to process measurements provided by a single receiver \cite{niu2019deep}.
Another study focuses on using MFP-based geoacoustic inversion methods to infer properties of the sea bottom, including the source's position in range and depth, by using a sparse receiver \cite{siderius1998broadband}. Both of these MFP approaches manage to do source localization but are challenged by the uncertainty in knowing the environment and the use of only a single receiver. To the best of our knowledge, all single hydrophone MFP approaches are designed for broadband finite-duration signals.

A class of popular approaches to underwater passive acoustic source localization makes use of \textit{waveguide invariant (WI)} theory. This theory describes range-dependent striation patterns charaterizing the spectrogram of a single hydrophone in a shallow water waveguide. The striation patterns are defined by the scalar WI parameter which can be obtained in a calibration step and makes it possible to extract range measurements \cite{cockrell2010robust}. WI-based ranging methods do not require much \textit{a priori} knowledge of the environment and are, therefore, an attractive alternative to MFP-based methods \cite{cockrell2010robust,rakotonarivo2012model,young2019waveguide,JanMey:C23}.
All aforementioned WI-based methods can provide the range of a source to a single acoustic receiver in different settings, but are unable to provide the source depth. In addition, WI-based methods are typically developed for specific source-receiver trajectories.

Single receiver source localization for sources emitting continuous signals based on cepstrum analysis has been presented in  \cite{trabattoni2023ship, dreo2022detection, trabattoni2020orienting}. Contrary to MFP- or WI-based approaches, this method relies on the extraction of TDOA measurements by means of cepstrum analysis.
Here, extracted \textit{time difference of arrival (TDOA)} measurements are deterministically associated with modeled TDOAs by relying on the amplitude sign to guide data association \cite{trabattoni2023ship, dreo2022detection, trabattoni2020orienting}. This deterministic data association approach ignores the potential presence of missed detections and false alarms and is thus prone to errors. 

\subsection{Contribution}

This paper aims to localize and track a single source using TDOA measurements extracted from the acoustic signal recorded by a single receiver.
Rather than performing data association deterministically, our method relies on probabilistic data association. It makes use of ray tracing and a statistical model to describe the propagation of signal paths through the water and the relation they have to the recorded measurements. A similar method for shallow water localization using vertical line array of receivers has been introduced in \cite{meyer2021probabilistic}, where a sequential Bayesian estimation method probabilistically associates \textit{direction of arrival (DOA)} measurements to modeled DOAs to jointly estimate the source's time-varying position in range and depth; for vertical arrays, the DOA is defined as the elevation angle. Our approach, referred to as \textit{probabilistic focalization}, has the ability to fuse TDOA measurements related to different propagation paths using probabilistic data association. Compared to MFP, probabilistic focalization can increase robustness to environmental uncertainty and model mismatch \cite{meyer2021probabilistic}. To the best of our knowledge, this paper introduces the first method that performs source localization using a single receiver in a shallow-water waveguide based on probabilistic data association.
By adapting the probabilistic focalization method to the considered TDOA measurement model, we aim to improve the robustness and accuracy of single receiver underwater source localization.

The contributions of this paper are summarized as\vspace{2mm} follows:
\begin{itemize}
\item We adapt the probabilistic focalization method \cite{meyer2021probabilistic} for our measurement model for TDOA data collection\vspace{3mm} \cite{dreo2022detection}.
\item We demonstrate improved estimation accuracy of our
methods with real-world collected\vspace{2mm} data.
\end{itemize}

\begin{figure}[t!]
\centering
\hspace{0mm}\mbox{\includegraphics[scale=0.33]{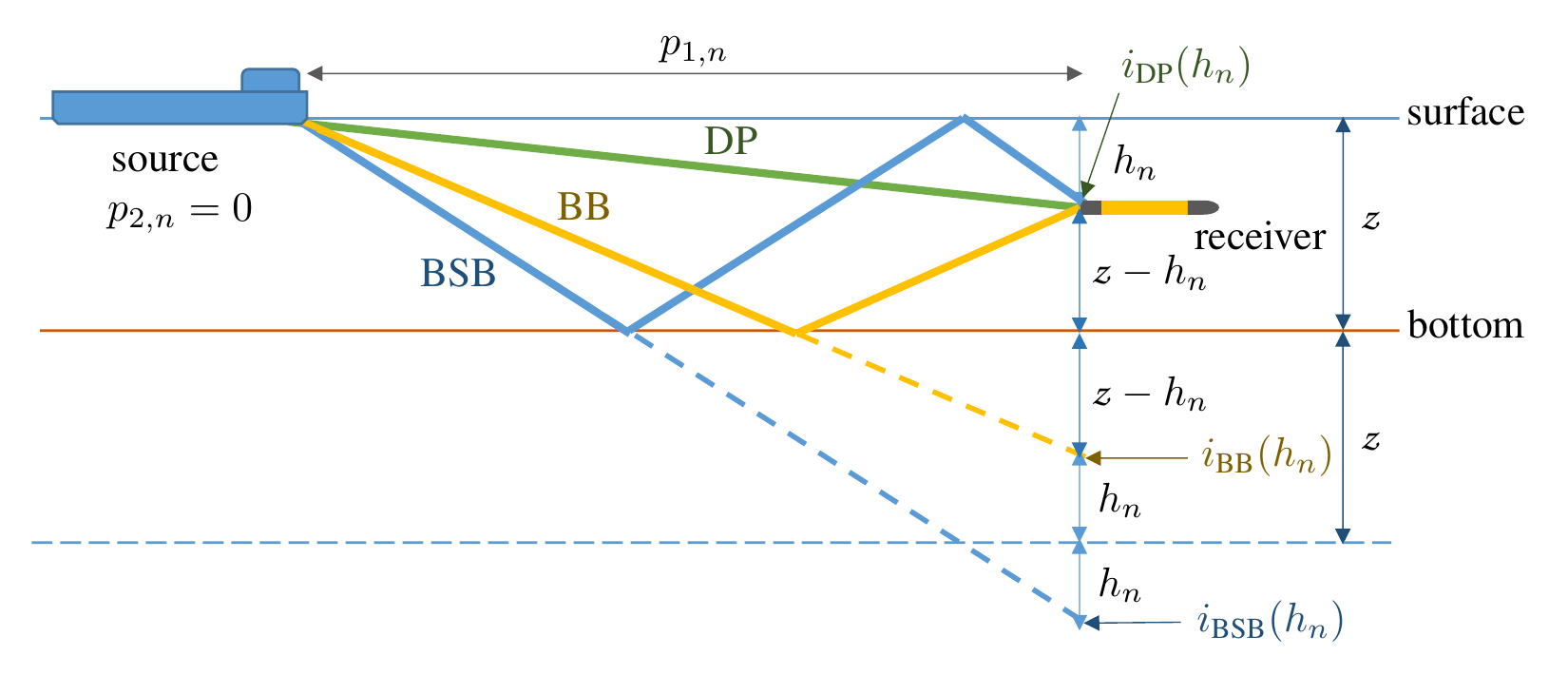}}
\vspace{-5.5mm}
\renewcommand{\baselinestretch}{1.05}\small\normalsize
\caption{Considered propagation paths used for TDOA-based localization of a single source using a single mobile receiver at time step $n$. Three isorays, \textit{direct path (DP)}, \textit{bottom bounce (BB)}, and \textit{bottom-surface bounce (BSB)}, and their imaged depths for a submerged mobile receiver at depth $h_n$ with a constant seafloor depth of $\mathpzc{z}$ are shown. The imaged depths are $i_{\text{DP}}(h_n) = h_n$, $i_{\text{BB}}(h_n) = 2\mathpzc{z} - h_n$, and $i_{\text{BSB}}(h_n) = 2\mathpzc{z} + h_n$.
}
\label{fig:imageRays}
\vspace{-5mm}
\end{figure}

For this paper, the focus is on an isoray environment model, where an iso-velocity SSP is assumed; this model is sufficient for the datasets used for performance evaluation. With this isorays model, the image method is used to calculate expected TDOAs \cite{1145559,HASSAB1976127}. The datasets are collected in a scenario with a surface source and a submerged mobile receiver. This scenario is illustrated in Fig. \ref{fig:imageRays}.

\section{Cepstrum Processing for TDOA Collection}
\label{sec:cepstrumProc}
\vspace{0mm}

In shallow water, the phase difference of different propagation paths generates constructive or destructive interference between the acoustic waves and modifies the received level of the wideband signal of the source. If the acoustic signal has a significant wideband component, multipath propagation leads to interference patterns in the form of striations in the spectrogram \cite{carey2009lloyd,kapolka2008equivalence}. The shape of these striations can encode the location of the acoustic source. Since interference is generated by the TDOA between the different paths and appears as oscillations in the log spectrogram, a natural way to analyze these oscillations is to perform the Fourier transform of the log spectrogram. The resulting signal is referred to as cepstrum \cite{bogert1963quefrency,oppenheim2004frequency,noll1964short,noll1967cepstrum}.

The cepstrum of a signal measured in a certain time window is defined as the inverse Fourier transform of its logarithmic spectrogram squared. It highlights periodicities of the measured spectrogram. The cepstrum output is in the quefrency domain whose unit is time. The resulting signal, however, is significantly different than the original time domain.
Similarly to the definition of spectrogram, the cepstrogram is obtained by computing the cepstrum of the signal over overlapped time windows,\vspace{-1mm} i.e.,
\begin{equation}
C(\tau,t)=\mathcal{F}^{-1} {\log(|S(f,t)|^2 )}. \nn
\vspace{1mm}
\end{equation}
Here, $\tau$ is the quefrency, $f$ is the frequency, $t$ defines the processed time window, and $|S(f,t)|$ is the spectrogram of the signal.
In this work, the cepstrogram is obtained from the spectrogram measured by a single receiver. A similar processing can also be performed for the beamformed signals recorded by an array of receivers \cite{dreo2022detection}.
The cepstrogram allows us to highlight periodicity in the log-spectrogram that can be either due to the harmonic structure of the source or the multipath propagation interferences.

Considering that the ship noise has a time-invariant harmonic structure whereas multipath interferences evolve with time as the ship moves through space, the cepstrogram can be decomposed as the sum of a source term and a propagation term by filtering based on the  \textit{Singular Value Decomposition (SVD)}. Here, the source term is the contribution of the singular vectors with the largest singular values while the propagation term is the contribution of the singular vectors with the lowest singular values
\cite{randall2017history,gao2013power,kruse2002new,ferguson2019improved}.
After SVD filtering and considering the propagation term, in the quefrency domain of the cepstrogram, each cosine component becomes a peak centered at the TDOA between the involved paths and whose amplitude is, at first approximation, the ratio between the arrival amplitudes, damped by its \textit{signal to noise ratio (SNR)}. This peak is inversely widened by the bandwidth of the signal, i.e. the frequency area where the SNR is sufficient to observe striations \cite{gao2013power,kruse2002new}.
The TDOA measurements, which are the inputs of our focusing algorithm, are obtained by extracting the local maxima of the modulus of the SVD-filtered cepstrogram \cite{kruse2002new}. 
To this end, we first compute an estimate of the background noise on the cepstrogram with a bi-dimensional median filter in the time-quefrency domain \cite{kruse2002new,trabattoni2020orienting} and then select only the local maxima with amplitudes larger than the local estimate of the background noise. 
After this processing step, a clustering algorithm is applied to filter out sparse peaks on the cepstrogram and select only the peaks which are close to each other in the time-quefrency domain. The applied filter is the \textit{density-based spatial clustering of applications with noise} algorithm \cite{dreo2022detection}. The TDOA measurements selected by this filter are the inputs of the focusing method described in this paper and are used to infer the distance and depth of the noise source. 

\section{Problem Formulation and System Model} \label{sec:sysModel}
\vspace{-.5mm}

Our goal is to estimate a time-varying mobile source state based on TDOA measurements of multiple propagation paths. At discrete time $n$, the source state is defined as $\V{x}_n \!\triangleq [\V{p}_n\trans \; v_n]\trans\rmv$, where $\V{p}_n = [p_{1,n}  \ist\ist p_{2,n}  ]^{\mathrm{T}}$ is the position in range and depth and $v_n$ is the source speed in range.  A fixed number of $K$ propagation paths is considered. This implies that there are  $L = {K \choose 2}$ pairs of propagation paths that can potentially generate TDOA measurements.  At every discrete time step $n$, each of the $L$ pairs of paths potentially contributes a TDOA measurement. As explained in Section \ref{sec:cepstrumProc}, cepstrum processing of the acoustic signal recorded by a single receiver provides $M_n$ TDOA measurements at each time $n$, denoted as $\V{z}_n = [z_{1,n} ... z_{M_n,n}]$. These measurements are subject to measurement-origin uncertainty and corrupted by missed detections and false alarms \cite{BarWilTia:B11}, i.e., $M_n$ can be larger, equal, or smaller than $L$.

Based on the statistical model to be introduced in this section, the \textit{minimum mean square error (MMSE)} estimator \cite{Kay:B93} is used to estimate the source state $\V{x}_n$, i.e.\vspace{.5mm},
\begin{equation}
\hat{\V{x}}^\text{MMSE}_{n} \,\triangleq\rmv \int\rmv \V{x}_n \ist f(\V{x}_n |\V{z}_{1:n}) \ist \mathrm{d}\V{x}_n \,,
\label{eq:mmse}
\vspace{.5mm}
\end{equation}
where $\V{z}_{1:n}$ is the vector of all TDOA measurements received up to time $n$.
This estimator requires the posterior \textit{probability density function (PDF)} $f(\V{x}_n |\V{z}_{1:n})$. This PDF, which is the probability of the source state given the measurements, is to be computed by the proposed probabilistic focalization approach.

\subsection{State Transition and Association Vectors}
\label{sec:vec_description}
%%%%%%%%%%%%%%%%%%%%%%%%%%%%%%%%%%%%%%%%%%%%%%%%%%%%%%%%%%%

The source state $\V{x}_n$ evolves over time according to  Markovian state dynamics, where $f(\V{x}_{n}|\V{x}_{n-1})$ is the state-transition PDF describing the motion of the source \cite{BarWilTia:B11}.

Each of the TDOA measurements $z_{m,n}$, $m \rmv\in\rmv \{1,\dots,M_n\}$ provided by cepstrum processing, as discussed in Section~\ref{sec:cepstrumProc}, is subject to measurement-origin uncertainty \cite{BarWilTia:B11}, i.e., there are missed detections, false alarms, and an unknown association between TDOA measurements and pairs of paths. Missed detections occur when a certain pair of propagation paths does not give rise to a TDOA measurement. To model missed detections, we define the probability that a pair provides a TDOA measurement, by $d_l\big(\V{p}_n \big) \!\in\rmv [0,1]$, $l \rmv\in\{1,$ $\dots,L\}$. These probabilities of detection are parameters to be set, which require some \textit{a priori} understanding of the propagation paths and their environment. If either path involved in any pair $l \rmv\in\{1,$ $\dots,L\}$ is not geometrically possible at the position $\V{p}_n $, $d_l\big(\V{p}_n \big) \! = 0$. A false alarm is a TDOA measurement that has not originated from any pair of propagation paths. Each of these measurements is not dependent on the state $\V{x}_n$, but is independent and identically distributed as $f_{\mathrm{FA}}\big( z_{m,n} \big)$. The PDF of false alarms is typically set to uniform on $[0,a_{\mathrm{FA}})$, where $a_{\mathrm{FA}}$ represents the maximum TDOA value within the distribution of false alarms. The value of $a_{\mathrm{FA}}$ can obtained by, assuming a minimum range, calculating the maximum path difference between the smallest and largest order of reflections, and dividing by the measured sound speed $c$. 
The number of false alarms follows a Poisson distribution with known mean $\mu_{\mathrm{FA}}$, which can be obtained by calculating the mean number of measurements and subtracting the expected value of true measurements, for each time step.

The unknown association between TDOA measurements and pairs of paths is modeled by a random \textit{data association vector} \cite{BarWilTia:B11}, that describes the associations between pairs of propagation paths $l \rmv\in\{1,$ $\dots,L\}$ and measurements $m \rmv\in\{1,$ $\dots,M_n\}$ at time $n$. Due to our $L$ pairs of paths, this vector is of dimension $L$, i.e., $\V{a}_n = \big[a_{1,n} \cdots\ist a_{L,n} \big]\trans\rmv\rmv$. The $l$th entry of $\V{a}_n$ indicates which measurement the $l$th pair generates or if there is a missed detection, i.e., $a_{l,n} \!=\rmv m \!\in\! \{1,\dots,M_n\}$ if the $l$th pair generates a TDOA measurement or $a_{l,n} \!=\rmv 0$ if the $l$th pair is missed. For a measurement $z_{m,n}$, $m \rmv\in\{1,$ $\dots,M_n\}$, if the index $m$ is not found in any of $\V{a}_n$'s elements, this means that no pair generated that measurement, indicating the presence of a false alarm.

\subsection{TDOA Measurement Model}
\label{sec:meas_likelihood} 
%%%%%%%%%%%%%%%%%%%%%%%%%%%%%%%%%%%%%%%%%%%%%%%%%%%%%%%%%%%

The generation of the detected TDOA measurement $z_{m,n}$ due to the pair of paths indexed by $l \rmv\in\{1,$ $\dots,L\}$ is modeled as
\begin{equation}
z_{m,n} = g_l(\V{p}_n) + w_{l,n},
\vspace{1.5mm}
\label{eq:measModel}
\end{equation}
where $w_{l,n}$ is zero mean Gaussian noise with variance $\sigma^2_{l}$. The model in \eqref{eq:measModel} establishes the relation between the pair of paths indexed by $l$ and the measurement indexed by $m$ at time $n$.

The modeled TDOA of the $l$th pair at position $\V{p}_n$ is the value $g_l(\V{p}_n)$. Due to the isorays model used in this paper, $g_l(\V{p}_n)$ is acquired using the geometry of the pair of paths and the image method. As seen in Fig.~\ref{fig:imageRays}, the model is able to calculate the length of each path at time step $n$, $q_k(\V{p}_n)$, $k \rmv\in\{1,$ $\dots,K\}$, by the Pythagorean theorem. For a submerged receiver at depth $h_n$, $i_k(h_n)$ is the imaged depth of the $k$th path, $p_{1,n}$ is the range between the source and receiver, and $p_{2,n}$ is the depth of the source, \vspace{.5mm}then 
\begin{equation}
q_k(\V{p}_n) = \sqrt{ \big(p_{2,n} - i_k(h_n) \big)^2 + p_{1,n}^{\ist 2}}. \nn
\vspace{1mm}
\end{equation}
The \textit{time of arrival (TOA)} $t_k(\V{p}_n)$ of each path is therefore $t_k(\V{p}_n) = q_k(\V{p}_n)/c$. The TDOA is then the difference of a pair of TOAs: $g_l(\V{p}_n) = |t_{l_1}(\V{p}_n) - t_{l_2}(\V{p}_n)|$, where $l_1$ and $l_2$ indicate the two paths involved in the $l$th pair.

The measurement model in \eqref{eq:measModel} describes the probability distribution of $z_{m,n}$ given the state $\V{x}_n$. Here, $l$ is the index of the considered pair of propagation paths. The conditional PDF resulting from \eqref{eq:measModel}, is therefore denoted as $f_l(z_{m,n} | \V{x}_n)$. With $\V{x}_n$ given, the value $g_l(\V{p}_n)$ is constant, meaning that $f_l(z_{m,n} | \V{x}_n)$ follows a Gaussian distribution with mean $g_l(\V{p}_n)$ and variance $\sigma^2_{l}$\vspace{1mm}.

\section{PDF Formulations for Estimation Method}
\label{sec:probFormEstimation}

The probabilistic focalization method provides an estimation method for the posterior PDF $f(\V{x}_n |\V{z}_{1:n})$. In this section, we will briefly discuss how it is used for our purposes of state estimation given TDOA measurements. The method first provides us the joint posterior PDF, which involves all source states, association variables, and TDOA measurements up to and including the current time $n$, i.e.\vspace{1.5mm},
\begin{align}
f(\V{x}_{1:n}, \V{a}_{1:n} | \V{z}_{1:n} ) &\propto f( \V{x}_{0}) \prod^{n}_{n'= 1} \rmv \psi(\V{a}_{n'}) f( \V{x}_{n'} | \V{x}_{n'\rmv-1}) \nn \\[1mm]
&\hspace{10.55mm}\times  \prod^{L}_{l= 1} r_l\big( \V{x}_{n'}, a_{l,n'}; \V{z}_{n'} \big) \hspace{.5mm}  \label{eq:factorOverall} \\[-6.5mm]
\nn %\\[-12mm]
\nn
\end{align} 
with
\begin{align}
 &r_l\big( \V{x}_{n}, a_{l,n}; \V{z}_{n} \big) \nn\\[4mm]
 &\hspace{4mm}\triangleq \begin{cases}
  \rmv\displaystyle \frac{d_{l}(\V{p}_n) \ist f_l\big(z_{a_{l,n},n} \big|\V{x}_n\big)}{\mu_{\mathrm{FA}} f_{\mathrm{FA}}\big( z_{a_{l,n},n} \big)} \ist, 
  &\!\rmv\rmv\rmv a_{l,n} \!\rmv\in\! \{1,\dots,M_n\} \\[.5mm]
  \rmv 1- d_{l}(\V{p}_n) \ist, & \!\rmv\rmv\rmv a_{l,n} \!=\rmv 0,
\end{cases} \nn\\[-5mm]
\nn
\end{align}
\vspace{2.5mm}and
\begin{align}
 \psi(\V{a}_n) &\triangleq \begin{cases}
  \rmv\displaystyle 0 \ist, 
  &\!\rmv\rmv\rmv \exists\, i,j \in [1,L] \,\text{s.t.}\, i \neq j, a_{i,n} = a_{j,n} \neq 0 \\[.5mm]
  \rmv 1 \ist, & \!\rmv\rmv\rmv \text{otherwise}.
\end{cases} \nn\\[-3.5mm]
\nn
\end{align}
The factor $r_l\big( \V{x}_{n}, a_{l,n}; M_{n} \big)$ provides the statistical model of how TDOA measurements are generated in the presence of measurement-origin uncertainty and $\psi(\V{a}_n)$ checks the validity of the association vector, not allowing more than one pair of paths to generate the same TDOA measurement.

With small variations, a detailed derivation of \eqref{eq:factorOverall} and its components is provided in \cite{meyer2021probabilistic,MeyKroWilLauHlaBraWin:J18}. Contrary to the probabilistic focalization model in \cite{meyer2021probabilistic}, more association vectors $\V{a}_n$ are present in our setting with TDOA measurements. With the DOA measurements, there is a necessity to keep the elements of $\V{a}_n$ ordered due to the DOAs of the propagation paths having a fixed order; for the TDOA measurements, there is no such restriction. In order for $\V{a}_n$ to be valid for TDOAs, the only requirement is that a measurement cannot be generated from more than one pair of paths, which is managed by $\psi(\V{a}_n)$. This is one of the subtle differences from the case of DOA measurements.

\begin{figure*}[t!]
\begin{minipage}[H!]{0.48\textwidth}
\centering

\begin{tikzpicture}
    \draw (0, 0) node[inner sep=0] {\includegraphics[scale=1]{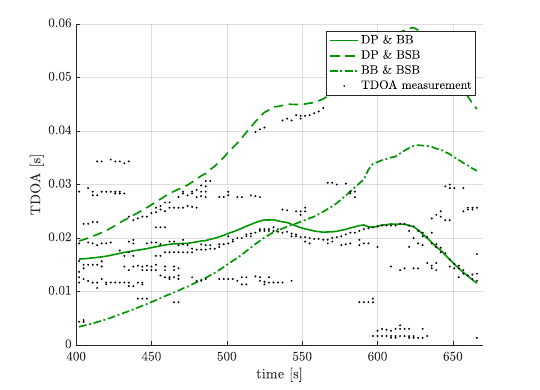}};
    \draw (.15, -3.43) node {\footnotesize $(a)$};
\end{tikzpicture}

\end{minipage}
\begin{minipage}[H!]{0.48\textwidth}
\centering

\begin{tikzpicture}
    \draw (0, 0) node[inner sep=0] {\includegraphics[scale=1]{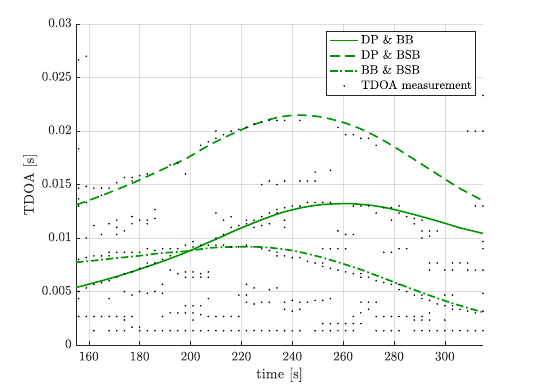}};
    \draw (.15, -3.43) node {\footnotesize $(b)$};
\end{tikzpicture}

\end{minipage}
\vspace{0mm}
\caption{TDOA measurements and modeled TDOA values for each pair of propagation paths for the Kraken (a) and Marfret (b) datasets. Modeled TDOA values are obtained following the discussion in Section \ref{sec:meas_likelihood}. }
\label{fig:data}
\vspace{-1mm}
\end{figure*}

\begin{figure*}[t!]

\begin{minipage}[H!]{0.48\textwidth}
\centering

\begin{tikzpicture}
    \draw (0, 0) node[inner sep=0] {\includegraphics[scale=1]{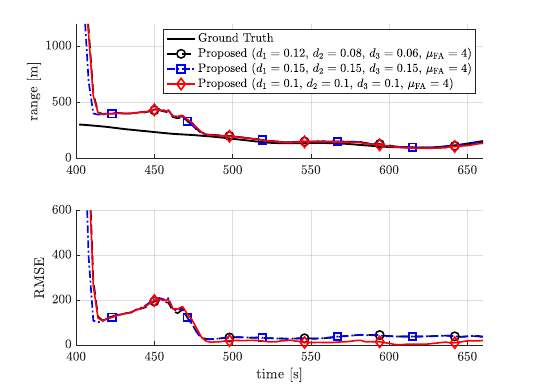}};
    \draw (.15, -3.43) node {\footnotesize $(a)$};
\end{tikzpicture}

\end{minipage}
\begin{minipage}[H!]{0.48\textwidth}
\centering

\begin{tikzpicture}
    \draw (0, 0) node[inner sep=0] {\includegraphics[scale=1]{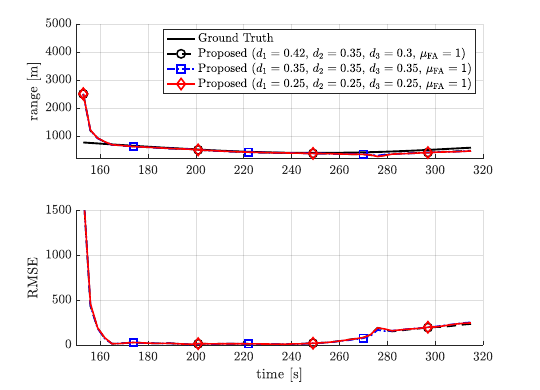}};
    \draw (.15, -3.43) node {\footnotesize $(b)$};
\end{tikzpicture}

\end{minipage}
\vspace{0mm}
\caption{Single receiver probabilistic focalization performance for the Kraken (a) and Marfret (b) datasets. Results are obtained by averaging over 50 runs of the proposed methods.}
\label{fig:results}
\vspace{-2mm}
\end{figure*}

As stated previously, the marginal posterior PDF $f(\V{x}_n |\V{z}_{1:n})$ is needed to compute the MMSE estimate of the source state over time. Based on the joint posterior $f(\V{x}_{1:n}, \V{a}_{1:n} | \V{z}_{1:n} )$, we compute accurate approximations of the marginal posterior $f(\V{x}_n |\V{z}_{1:n})$ by using a particle-based implementation of the \textit{sum-product algorithm} \cite{KscFreLoe:01,IhlFisMosWil:05,MeyBraWilHla:J17}. This type of processing  leads to a set of particles $\big\{(\V{x}_n^{(j)},w_n^{(j)})\big\}_{j=1}^J$, where $J$ is the number of particles, that represents the posterior $f(\V{x}_n |\V{z}_{1:n})$ at each time $n$. A particle-approximation of the MMSE estimate $\hat{\V{x}}^\text{MMSE}_{n}$ in \eqref{eq:mmse} can finally be\vspace{-1mm} obtained as 
\begin{equation}
\hat{\V{x}}^\text{MMSE}_{n} \ist=\ist \sum_{j=1}^{J} \omega_{n}^{(j)} \V{x}_{n}^{(j)}. \nn
\vspace{-1mm}
\end{equation}
Next, the proposed method is evaluated using two real datasets collected by \textit{NATO STO Centre for Maritime Research and Experimentation}\vspace{-2mm}.

\section{Performance Evaluation}
\vspace{-.5mm}
\label{sec:results}

In this section, we will analyze the performance of our adapted probabilistic focalization method for our measurement model for TDOA data collection. The datasets used to evaluate this performance were recorded off the Italian coast in front of the Cinque Terre coastline with a depth varying between 50 and 350 m; a detailed description of the experiments and their datasets is provided in \cite{dreo2022detection}. The two datasets both involve a source moving along the sea surface and a moving mobile receiver. This mobile receiver was equipped with a GPS to record the ground truth data accurately. The mean sound speed for both situations was about 1508 m/s, where isovelocity is assumed. We here focus on only one section of each of the datasets, specifically the section that provides the most useful data for our analysis. In the other sections, the SNR is too low, and consequently, there are too many missed detections for a successful localization and tracking of the source.

The first dataset's source is a small workboat, named \textit{Kraken}, that remains within 500 m of the receiver with a closest point of approach of about 100 m, at a speed of about 5-6 kn, or 2.6-3.1 m/s. The mobile receiver had a speed of about 0.37 m/s with a maximum depth of roughly 40 m and a minimum depth of about 5 m. The depth of the sea floor is recorded to be a constant 65 m in the area of interest. The second dataset's source is a 125-m long cargo ship, named \textit{Marfret Niolon}, that has a farther distance from the receiver, with a maximum range of 3 km and a closest point of approach of about 500 m, at a speed of about 12-13 kn, or 6.1-6.7 m/s. The mobile receiver had a speed of 0.38 m/s with a depth between about 7-30 m. The sea floor was at depth 73 m.

The setup for the implementation of the estimation method is as follows. The area of interest for the source is for a range of [0 m, 5000 m] from the receiver and a depth of 0 m from the sea surface. The surveillance area could cover a range of depths, but since this experiment focuses on surface sources, we limit depth to a single value.  For the source state $\V{x}_n = [\V{p}_n\trans \; v_n]\trans\rmv$, the motion is modeled\vspace{0mm} as
\begin{equation}
\V{x}_{n} =  { \begin{pmatrix}
   1 & 0 & T_n  \\
   0 & 1 & 0  \\
   0  & 0  & 1 
  \end{pmatrix} } \ist\V{x}_{n-1} + { \begin{pmatrix}
   \ist\frac{T_n^2}{2} & 0 \\
   0  & T_n  \\
   T_n & 0 
  \end{pmatrix} } \V{u}_{n},
  \label{eq:stateTransition}
  \vspace{2mm}
\end{equation}
where the scan time $T_n = 3$s and the driving noise $\V{u}_n \sim \Set{N}(\V{0},\M{\Sigma}_u )$ with $\M{\Sigma}_u  \!= \mathrm{diag} \big\{0.05 \ist\ist \text{m}^2/\text{s}^4 \hspace{2.5mm} 0 \ist\ist \text{m}^2/\text{s}^2 \big\}$. The two-dimensional Gaussian random vectors $\V{u}_n$ are independent and identically distributed. There is no driving noise in depth since we are estimating for a surface source with a constant depth of 0m. With this source motion model, we now have a definition for the state transition function $f(\V{x}_{n}|\V{x}_{n-1})$, as described in Section \ref{sec:vec_description}. The prior distribution follows $f(\V{x}_0) = f(\V{p}_0)f(v_0)$, where $f(\V{p}_0)$ is uniform over the surveillance area and $f(v_0)$ is zero-mean Gaussian with a standard deviation of 5 m/s, hereby giving us the factor found in (\ref{eq:factorOverall}). For the particle-based implementation, $J = 10^4$ particles are used.

For geometrical reasons (cf.~Fig.~\ref{fig:imageRays}), in this shallow water environment, $K = 3$ paths are sufficient for the estimation of the source's position. Since $K = 3$, we have $L = 3$ pairs of paths, i.e., \textit{direct path (DP)} and \textit{bottom bounce (BB)} ($l=1$), DP and \textit{bottom-surface bounce (BSB)} ($l=2$), and BB and BSB ($l=3$). The noise standard deviation for the TDOA measurements defined in \eqref{eq:measModel} is $\sigma_l = 5 \times 10^{-4}$ s for $l\rmv\in \{1,2,3\}$. The value $a_{\mathrm{FA}}$ was found to be about 0.1 s for both datasets.

\vspace{0mm}
Three parameter settings for each dataset are evaluated using the proposed estimation method. All probabilities of detections are independent of the source position, i.e., $d_l = d_l(\V{p}_n) $, $l \in \{1,2,3\}$. For the Kraken dataset, we use the following three sets of parameters: (i) $d_1 = 0.12$, $d_2 = 0.08$, $d_3 = 0.06$, and $\mu_{\mathrm{FA}} = 4$; (ii) $d_1 = 0.15$, $d_2 = 0.15$, $d_3 = 0.15$, and $\mu_{\mathrm{FA}} = 4$; and (iii) $d_1 = 0.1$, $d_2 = 0.1$, $d_3 = 0.1$, and $\mu_{\mathrm{FA}} = 4$. In addition, for the Marfret dataset, the three sets of parameters are: (i) $d_1 = 0.42$, $d_2 = 0.35$, $d_3 = 0.3$, and $\mu_{\mathrm{FA}} = 1$; (ii) $d_1 = 0.35$, $d_2 = 0.35$, $d_3 = 0.35$, and $\mu_{\mathrm{FA}} = 1$; (iii) and $d_1 = 0.25$, $d_2 = 0.25$, $d_3 = 0.25$, and $\mu_{\mathrm{FA}} = 1$\vspace{-.5mm}.

In Fig.~\ref{fig:data}, the TDOA measurements are provided for each of the two datasets, Kraken and Marfret, in addition to the modeled TDOA values for each of the pairs of paths, as discussed in Section \ref{sec:meas_likelihood}. Fig.~\ref{fig:results} shows our final results of probabilistic focalization for both datasets for 50 trials. We inspect the \textit{root mean-squared-error (RMSE)} as an error metric for the position estimate at each time step, in addition to the mean range estimate, across the 50 trials. A depth estimate is not provided here as it is known that the source is on the surface, i.e., at an assumed depth of 0 m. Due to the large surveillance range, the initial range estimate is 2500 m, which leads to a rather large initial RMSE value. The estimate is able to calibrate and achieve a low RMSE within just a few seconds. 

In the Kraken scenario, the proposed method is able to maintain low estimation error. In the Marfret scenario, the localization error is low initially but increases after time 250 s due to a combination of a high number of missed detections and an increased  false alarm rate. The different settings illustrate the importance of parameter selection and how all are able to reliably localize the source with varying degrees of accuracy\vspace{-1mm}.

\section{Conclusion}
\label{sec:conclusion}

In the paper, we adapted the probabilistic focalization \cite{meyer2021probabilistic} to a scenario with a single mobile underwater receiver. TDOA measurements are extracted from the acoustic signal of a single receiver through cepstrum analysis. By formalizing the probabilistic focalization approach for a TDOA measurement model, we develop an algorithm that probabilistically associates observed TDOAs with modeled TDOAs in the presence of unwanted false alarms and missed detections. Our results based on real data, demonstrate accurate source localization based on datasets analyzed in \cite{dreo2022detection}. Promising future research venues are the combination of probabilistic focalization with neural networks \cite{LiaMey:J24} and particle flow techniques\cite{ZhaMey:J24}. 
\vspace{.5mm}

\section{Acknowledgement}
This work was supported in part by the Office of Naval Research under Grants  N00014-23-1-2284 and N00014-24-1-2021 and in part by the NATO Allied Command Transformation (ACT) under the CMRE Autonomous Anti-Submarine Warfare research program.
 
\vspace{.5mm}

\renewcommand{\baselinestretch}{1}
\selectfont

\bibliographystyle{IEEEtran}
\bibliography{IEEEabrv,Temp,Books,Papers}

\end{document}